\documentclass[amsmath,amssymb,twocolumn]{revtex4-1}
\usepackage{graphicx}
\usepackage{hyperref}
\usepackage{url}
\def\registered{{\ooalign{\hfil\raise .00ex\hbox{\tiny R}\hfil\crcr\mathhexbox20D}}}
      \begin{document}
      \title{Deep phase modulation interferometry}
\author{Gerhard Heinzel} 		\email[Corresponding author:]{gerhard.heinzel@aei.mpg.de}
\author{Felipe Guzm\'an Cervantes} 	\email[Corresponding author:]{felipe.guzman@aei.mpg.de}\email[\\]{felipe.guzman@nasa.gov}
\author{Antonio F. Garc\'ia Mar\'in}
\author{Joachim Kullmann}
\author{Karsten Danzmann}
\affiliation{Albert-Einstein-Institut Hannover (Max-Planck-Institut f\"ur Gravitationsphysik, and Leibniz Universit\"at Hannover), Callinstra\ss e 38, 30167 Hannover, Germany}
\affiliation{$^{\dagger}$NASA Goddard Space Flight Center, 8800 Greenbelt Road, Greenbelt, MD 20771, USA}
\author{Wang Feng}
\affiliation{Purple Mountain Observatory, CAS, 2 West Beijing Road, Nanjing 210008, China}

      \begin{abstract}
We have developed a method to equip homodyne interferometers with the capability
to operate with constant high sensitivity over many fringes for continuous
real-time tracking. The method can be
considered as an extension of the ``$J_1\dots J_4$'' methods, and its
enhancement to deliver very
sensitive angular measurements through Differential Wavefront Sensing is
straightforward. Beam generation requires a sinusoidal phase modulation of
several radians in one interferometer arm. On a stable optical bench, we have
demonstrated a long-term sensitivity over thousands of seconds of
0.1\,mrad$/\sqrt{\rm Hz}$ that correspond to 20\,pm$/\sqrt{\rm Hz}$ in length,
and 10\,nrad$/\sqrt{\rm Hz}$ in angle at millihertz frequencies.
      \end{abstract}
\maketitle
      
      \section{Introduction}
      \label{intro}%
Optical interferometers with sub-wavelength resolution are useful in many
optical metrology
applications, such as, for example, length measurements, gravitational wave
detection, 
wavefront sensing, and surface profiling, among others. Our technique was
developed in the context of continuously measuring the position
and orientation of a free-floating test mass for space-based gravitational wave
detection\,\cite{anza}, although the method is useful for other applications as
well. 
Other techniques for
the optical readout of free-floating
test masses at millihertz frequencies are currently under investigation, such as
a polarizing heterodyne interferometer
reaching a sensitivity of about
$300\,\mathrm{pm}/\sqrt{\mathrm{Hz}}$\,\cite{oro_berlin}, a compact 
homodyne interferometer with a sensitivity of 
$100\,\mathrm{pm}/\sqrt{\mathrm{Hz}}$\,\cite{oro_bmh}, and a robust 
implementation of an optical lever with a readout noise level of 
$100\,\mathrm{pm}/\sqrt{\mathrm{Hz}}$\,\cite{oro_napoli}. Another method to do
this is heterodyne interferometry as developed for LISA Pathfinder\,\cite{ltp}
with a sensitivity of better than
$5\,\mathrm{pm}/\sqrt{\mathrm{Hz}}$\,\cite{ltpsubtraction}.
The method we present here achieves an optical pathlength measurement
sensitivity of the order of
$20\,\mathrm{pm}/\sqrt{\mathrm{Hz}}$, and with an angular resolution better than
$10\,\mathrm{nrad}/\sqrt{\mathrm{Hz}}$,
both above 3\,mHz. The conversion from real test mass motion to optical
pathlength is given by the
interferometer topology, and is in our case about a factor of 2, which yields a
test mass motion resolution of approximately
$10\,\mathrm{pm}/\sqrt{\mathrm{Hz}}$.
Those interferometers with the highest accuracy, namely Fabry-Perot
interferometers on
resonance or recycled Michelson interferometers on a dark fringe\,\cite{geo},
have
a dynamic range of a small fraction of one fringe only. High
resolution and wide dynamic range can be simultaneously achieved by, e.g.,
active feedback or heterodyning, each of which has disadvantages. Active
feedback transfers the inherent non-linearity of the feedback actuator to the
output signal or requires another stabilized laser and a measurement of the
high-frequency beat note. Heterodyning, on the other hand,
requires a complex setup to generate the two coherent beams with a constant
frequency difference, typically involving two acousto-optic modulators (AOMs)
with associated frequency generation and RF power amplification.
Other methods to overcome these limitations involve variations of sinusoidal
phase shifting
interferometry\,\cite{sasaki_1986,sasaki_1987,deGroot_2008,deGroot_2009,
Falaggis_2009}, reporting accuracies of the order of 1\,nm. These methods are
typically used in ``single-shot'' mode for static applications such as surface
profiling, whereas our method is designed for continuous real-time, long-term
tracking of a moving target with low noise at millihertz frequencies.
In particular, the so-called ``$J_1\dots J_4$'' method\,\cite{su89,jin91,su93}, 
involves a sinusoidal phase modulation at a
fixed frequency $f_{\rm mod}$ with modulation depths $m\approx1\dots5$ in one
arm of the interferometer. The spectrum of the resulting photocurrent has
components  at integer multiples of $f_{\rm mod}$, with amplitudes that can be
written in terms of the Bessel functions $J_n(m)$ (hence the name) and the phase
difference $\varphi$ due to the optical pathlength difference. The methods then
proceed to use analytical formulae to solve for the unknowns $m$ and $\varphi$,
after obtaining the harmonic amplitudes from a spectrum analyzer or a Fast
Fourier Transform (FFT) of the digitized time series. The accuracies reported
are of order 10\dots100\,mrad (1.7\,\dots17\,nm) for a laser wavelength of
1064\,nm. We
generalize this approach by using a higher modulation index $m$ (up to 10 or
20) and making use of all harmonics up to an order $N\approx m$. These are more
observations than the four unknowns ($m$, $\varphi$, modulation phase and a
common factor), making an analytical solution impossible. Instead we use a
numerical least-squares solution which allows consistency checks and improves
the signal-to-noise ratio. For our typical applications we keep $m$ near
constant at an optimal value and take $\varphi$ as useful output, achieving an
accuracy of better than 0.12\,mrad$/\sqrt{\rm Hz}$ that correspond to
20\,pm$/\sqrt{\rm Hz}$ in length, and
10\,nrad$/\sqrt{\rm Hz}$ in angle at millihertz frequencies.
As compared to heterodyne interferometers, more complex data processing is
necessary to recover the optical pathlength from the measured photocurrent.
However,
with the availability of inexpensive processing power, this
computational complexity is often preferable to additional optics and
electronics
hardware needed for the optical heterodyning.

\section{Theory}
\label{chap_dpm_theory}
\begin{figure*}[bth]
\begin{center}
\includegraphics[width=\textwidth]{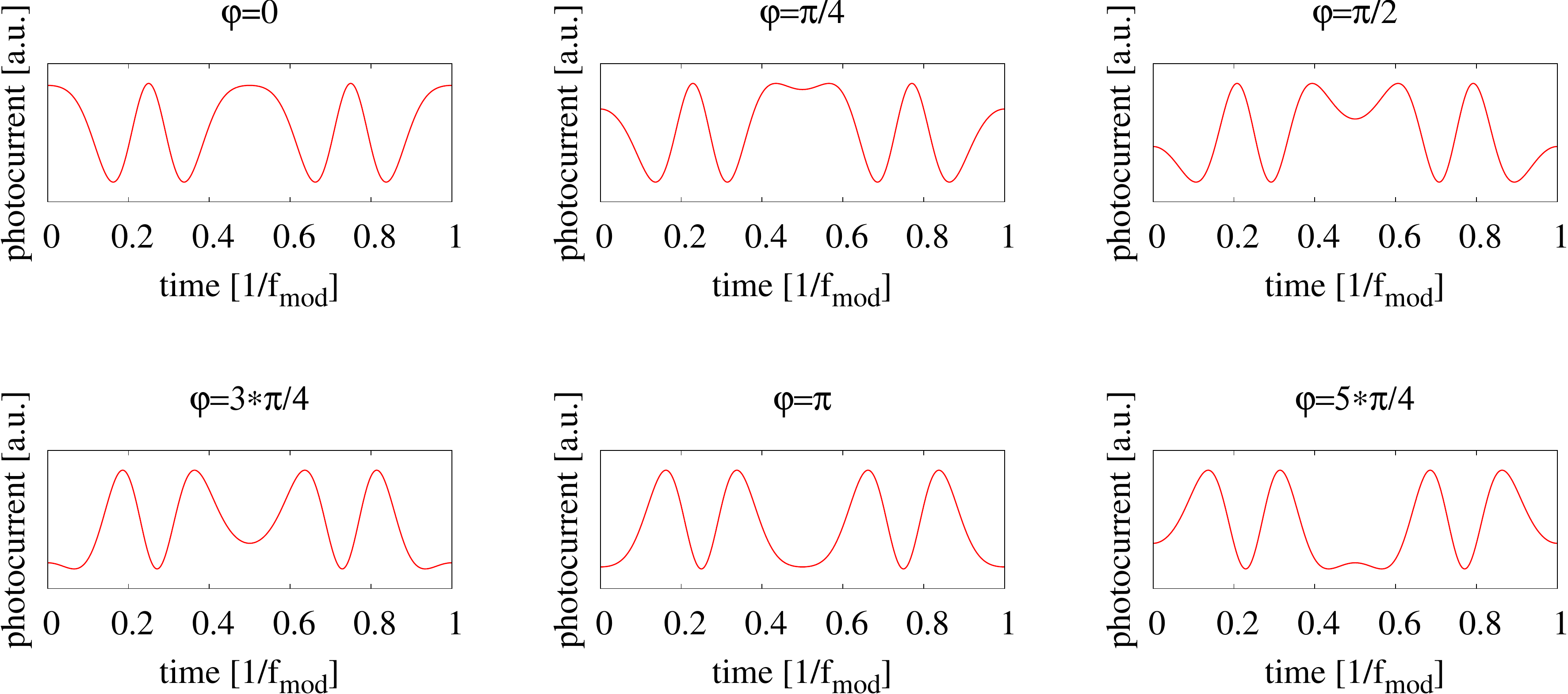}
\caption{Waveform of the obtained interferogram for different operating points
of the interferometer phase $\varphi$ with a modulation depth
$m=6\,\mathrm{rad}$.}
\label{oro_waveform}
\end{center}
\end{figure*}

\begin{figure}[bth]
\begin{center}
\includegraphics[width=\columnwidth]{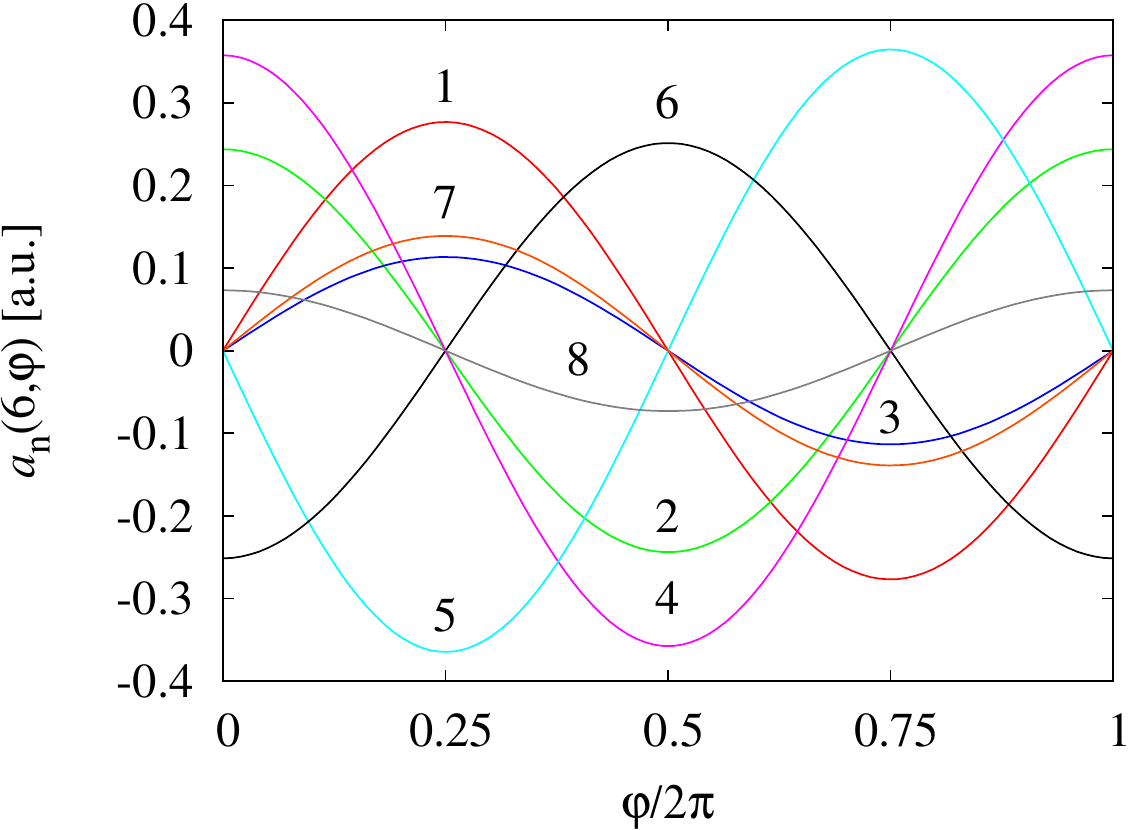}
\caption{Dependence of the harmonics amplitudes $a_n(m,\varphi)$ with respect to
the interferometer phase $\varphi$ with a modulation depth $m=6\,\mathrm{rad}$.}
\label{harmonics}
\end{center}
\end{figure}

The signal $V_{\mathrm{PD}}(t)$ of a photodetector at the output of a
phase-modulated homodyne interferometer can be 
expressed as
\begin{equation}
V_{\mathrm{PD}}(t)=A\,\left[\,1+C\,\mathrm{cos}\left(\varphi+m\,\mathrm{cos}
\left(\omega_{\mathrm{m}}\,t+\psi \right)\, 
\right)\,  \right],
\label{dpm_pdsignal}
\end{equation}
where $\varphi$ is the interferometer phase, $m$ is the modulation depth,
$\omega_{\mathrm{m}}=2\pi\,f_{\mathrm{m}}$ is 
the modulation frequency, $\psi$ is the modulation phase, $C\leq1$ is the
contrast, and $A$ combines nominally constant factors
such as light powers and photodiode efficiencies.
The interferometer output is periodic with $f_{\mathrm{m}}$ and its signal
waveform characteristically depends on the 
interferometer phase $\varphi$. Figure~\ref{oro_waveform} illustrates typical
waveforms obtained for various states of 
$\varphi$.
The expression of Equation~\ref{dpm_pdsignal} can be expanded into its harmonic
components as:
\begin{equation}
V_{\rm PD}(t)=V_{\rm
DC}(\varphi)+\sum\limits_{{n}=1}^{\infty}a_{n}(m,\varphi)\cos ({n}(\omega_{\rm
m}t+\psi))
\label{bessel_exp}
\end{equation}
with
\begin{eqnarray}
a_{n}(m,\varphi)&=&k\,J_{n}(m)\,\cos\left(\varphi+n\frac{\pi}{2} \right)
,\,\,\mathrm{and}\\
V_{\rm DC}(\varphi)&=&A\left(1+C\,J_0(m)\cos\varphi\right),
\label{dpm_an}
\end{eqnarray}
where $k=2CA$, and $J_\mathrm{n}(m)$ are the Bessel functions.
Figure~\ref{harmonics} shows the dependence of the 
harmonic amplitudes $a_n(m,\varphi)$ in terms of $\varphi$. Our technique is
centered around these harmonic amplitudes $a_n(m,\varphi)$ which on the one hand
can be directly measured by numerical Fourier analysis of the photocurrent, and
on the other hand have the above analytical relationships to the unknowns
$\varphi$, $m$, $\psi$, $k$.
The technique we present here uses higher
modulation depths $m \ge 6$ to set up an overdimensioned 
system of equations that can be numerically solved for the four sought
parameters $\varphi$, $m$, $\psi$, and the common 
factor $k$ by a least-squares fit algorithm. The information of the harmonic
amplitude $a_0(m,\varphi)$, 
corresponding to the DC component $V_{\rm DC}(\varphi)$ is not used by the
fit algorithm, since it usually contains a 
higher noise level due to large variations in environmental and equipment
conditions such as room illumination and 
electronic noise, among others. However, it is useful for computation of the
interferometer visibility and alignment signals.
\section{Data processing}\label{dpm_fit}
\begin{figure}[bth]
\begin{center}
\includegraphics[width=\columnwidth]{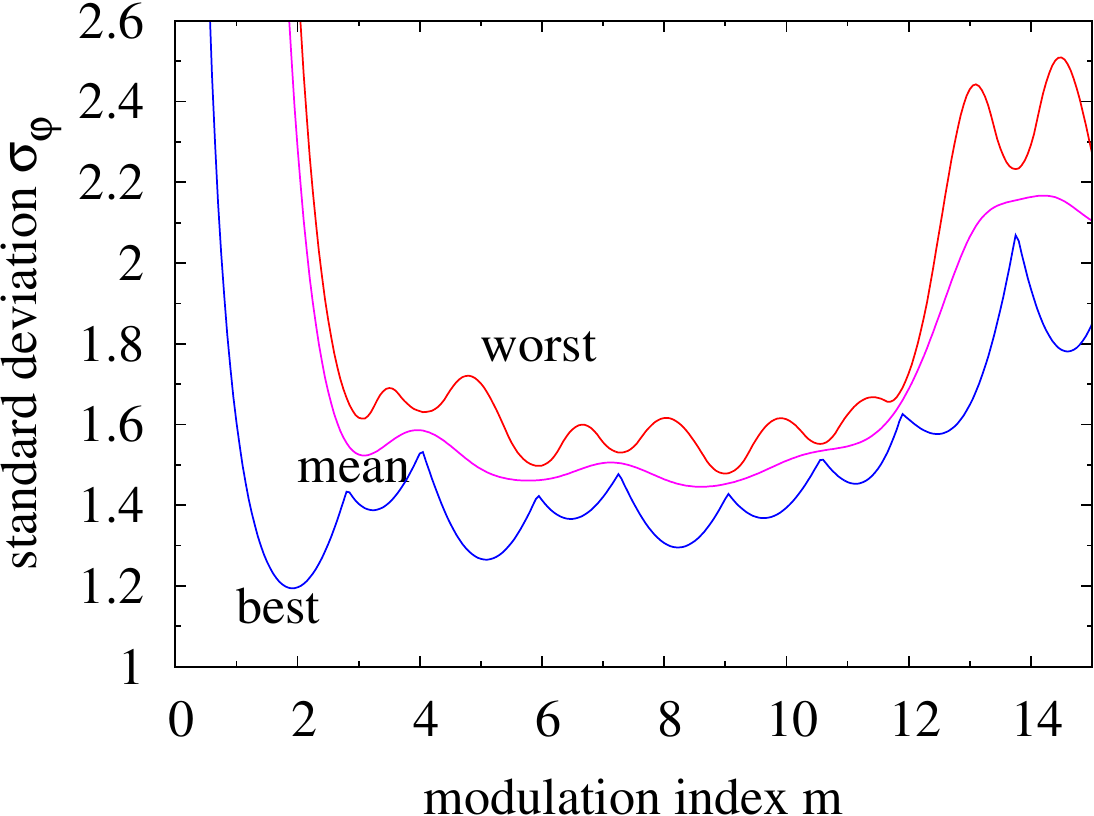}
\caption{\label{distb}Ideal resolution in $\varphi$ as function of the
modulation index $m$ for $N=10$, for the best and worst $\varphi$ as well as the
average for all $\varphi\in[0,2\pi]$.}
\end{center}
\end{figure}

\begin{figure}[bth]
\begin{center}
\includegraphics[width=\columnwidth]{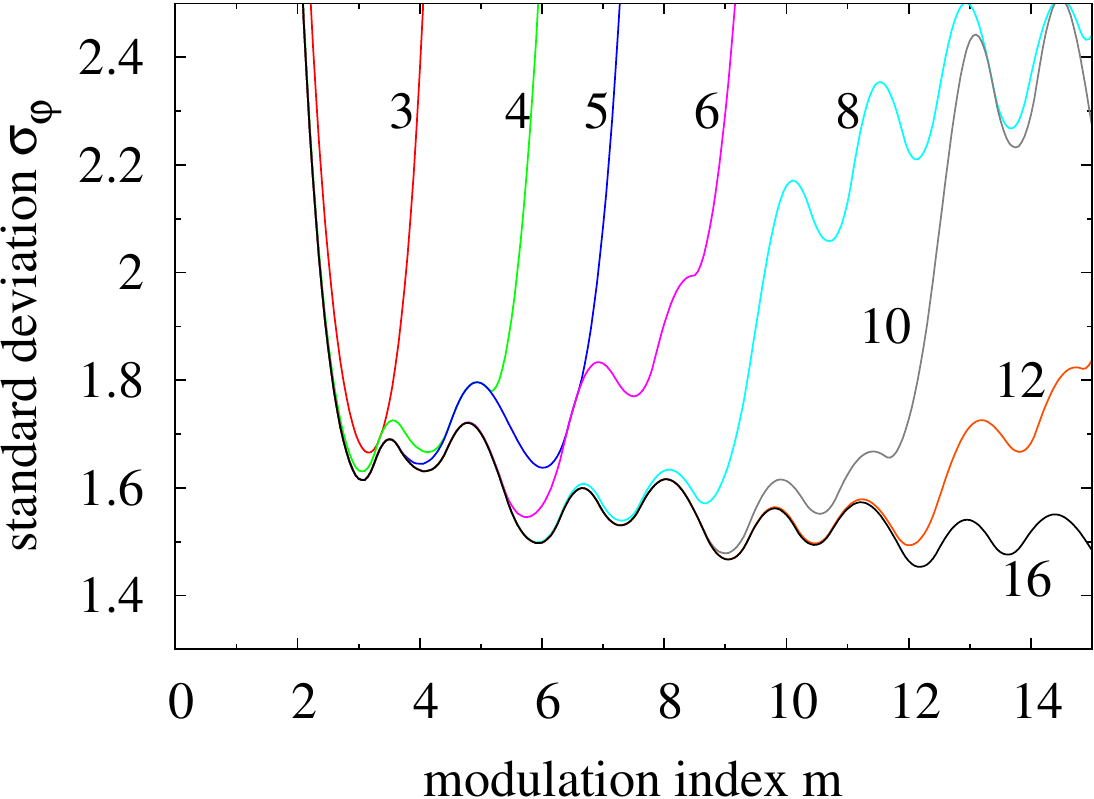}
\caption{\label{dista}Ideal resolution in $\varphi$ as function of the
modulation index $m$ for different orders $N$, for the worst value of $\varphi$
at each point of each curve.}
\end{center}
\end{figure}

\begin{figure}[bth]
\begin{center}
\includegraphics[width=\columnwidth]{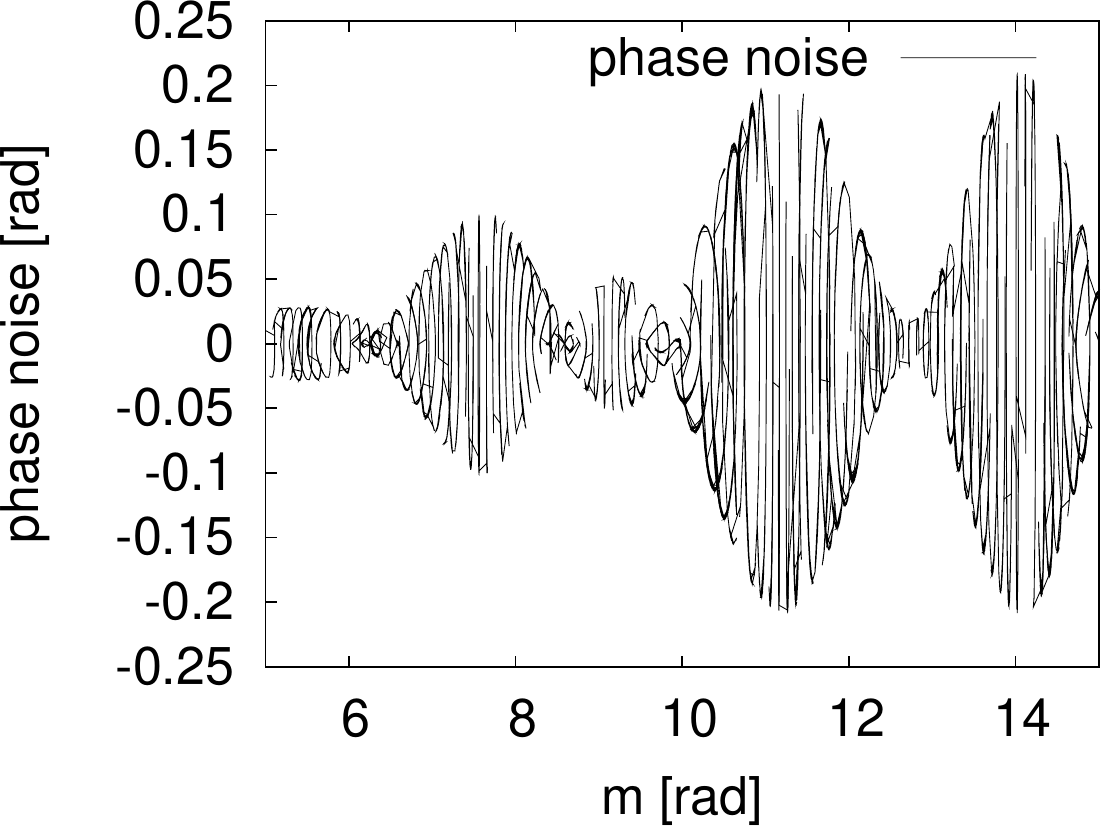}
\caption{Dependence of the measured phase noise with $m$.}
\label{dpm_mvariable}
\end{center}
\end{figure}

The signal $V_{\rm PD}(t)$ measured at the photodetector is digitized after
appropriate analog processing and anti-alias filtering. The sampling rate
$f_{\rm samp}$ is arranged to be coherent to the
modulation frequency $f_{\rm mod}$. The time series is split in segments of
length $N_{\mathrm{FFT}}$ samples that are 
processed by a Discrete Fourier Transform in order to compute
$N=N_{\mathrm{FFT}}/2$ complex amplitudes 
$\tilde{\alpha}_n(m,\varphi)$. A non-linear fit algorithm is applied to match
the measured $\tilde{\alpha}_n(m,\varphi)$ to the complex 
amplitudes $c_n$ computed from the model
\begin{equation}
\alpha_{n}(m,\varphi)=a_n(m,\varphi)\,e^{{\rm i}n\psi}.
\label{dpm_cn}
\end{equation}
There is a total of $2N$ equations that can be set up in two uncorrelated system
of equations:
\begin{eqnarray}
n\psi&=&\arctan\left(
\frac{\Im\{\alpha_n(m,\varphi)\}}{\Re\{\alpha_n(m,\varphi)\}}\right),n=1,2
,3\dots N,\,\,\,\,\,\,\,\,\\
a_n(m,\varphi)&=&\alpha_n(m,\varphi)\,e^{{\rm -i}n\psi},\,\,n=1,\,2,\,3\dots N,
\label{dpm_syseqn}
\end{eqnarray}
where $\alpha_n(m,\varphi)\,e^{{\rm -i}n\psi}$ is a real number.  For the
measured 
$\tilde{\alpha}_n(m,\varphi)\,e^{{\rm -i}n\psi}$, this is not exactly the case
due to noise and phase distortions introduced by the analog 
electronics. In order to solve the system of equations, a Levenberg-Marquardt
fit algorithm\,\cite{marquardt, numrec} is applied to
minimize the least-squares expression
\begin{equation}
\chi^2=\sum\limits_{n=1}^{N}{\left(\alpha_n(m,\varphi)-\tilde{\alpha}_n(m,
\varphi)\right)}^2,
\label{dpm_chi}
\end{equation}
where $\chi^2$ is a four dimensional function of $m$, $\varphi$, $\psi$, and
$k$. In practice, these parameters barely vary 
between consecutive segments of length $N_{\mathrm{FFT}}$, giving good starting
values for a rapid convergence of the fit.
Only in the case this is not accomplished such as upon initialization or after
large disturbances, a modified version of the more robust 
Nelder-Mead Simplex algorithm\,\cite{nelder} is applied as initial step.
In order to find best values of the modulation index $m$ and the number of bins
$N$ for optimum performance, we conducted a 
numerical analysis of the $4\times 4$ Hessian matrix of $\chi^2$ that is given
by
\begin{equation}
H=(H_{ij})=\left(\frac{\partial^2\chi^2}{\partial \Omega_i\partial
\Omega_j}\right),
\end{equation}
where $\Omega=\{m,\varphi,\psi,k\}$ are the four parameters. The inverse of the
Hessian matrix $H^{-1}=(\eta_{ij}$) yields 
information about the parameter estimates on the variances $\sigma^2$ and
correlation coefficients $\rho_{ij}$:
\begin{eqnarray}
\sigma^2_{\Omega_i} &\propto& \eta_{ii},\\
\rho_{ij} &=& \frac{\eta_{ij}}{\sqrt{\eta_{ii}}\sqrt{\eta_{jj}}}.
\end{eqnarray}
An excursion of $\varphi$ over the range $[0,2\pi]$ --which corresponds to one
interferometer fringe-- was conducted in 64 steps 
by fixed $N$ and $m$ in order to compute the best, worst and average values of
the standard deviation 
$\sigma_{\Omega_i}(N,m,\varphi)$, which are shown in Figure~\ref{distb}.
Assuming worst case values of the variances 
\begin{equation}
\widehat{\sigma}^2_{\Omega_i}(N,m) = \max_{\varphi\in[0,2\pi]}
\sigma^2_{\Omega_i}(N,m,\varphi),
\end{equation}
we run a similar analysis varying $N$ and $m$ to evaluate for best resolution of
any value of $\varphi$, which is our main 
measurement and often not entirely under control. The results are shown in
Figure~\ref{dista}.
This analysis revealed useful parameter estimates for $3 \leq m \leq N$, and
possible best values of $m$ for
minimum $\sigma_{\varphi}$ in the cases of $m=6, N\ge8$ and $m=9, N\ge10$,
suggesting best 
resolutions of $\varphi$. These results are only rough guidelines, since real 
instrument noise has not been yet considered. The dominant noise sources have,
however, been investigated experimentally, as discussed in
Section~\ref{noise_inv} below. In addition, software simulations
of the fit routine were run with synthetic data as 
input. Hardware characteristics of the data acquisition system (DAQ) such as
digitization effects and frequency response of the 
anti-aliasing filter were considered in the generation of mock-data, using
Equation~\ref{dpm_pdsignal} as nominal noise-free 
model. We introduced two independent mock-data sets into two virtual 
DAQ channels, by linearly increasing $m$ with $N=10$ bins, and recorded the fit
output (phase $\varphi$).
We then computed their phase difference and extracted the nominal offset in oder
to obtain the dependence of the 
phase fluctuations (noise) with $m$, which is shown in
Figure~\ref{dpm_mvariable}.
A minimum can be observed around $m=9.5-10$, which is consistent with the
analysis presented in Figure~\ref{dista}. Hence, a
DAQ test system for real optical length measurements was set up with $N=10$
bins, and a modulation index $m\approx9.7$.
\section{Experimental setup}\label{exp_setup}
\begin{figure*}[bth]
\begin{center}
\includegraphics[width=\textwidth]{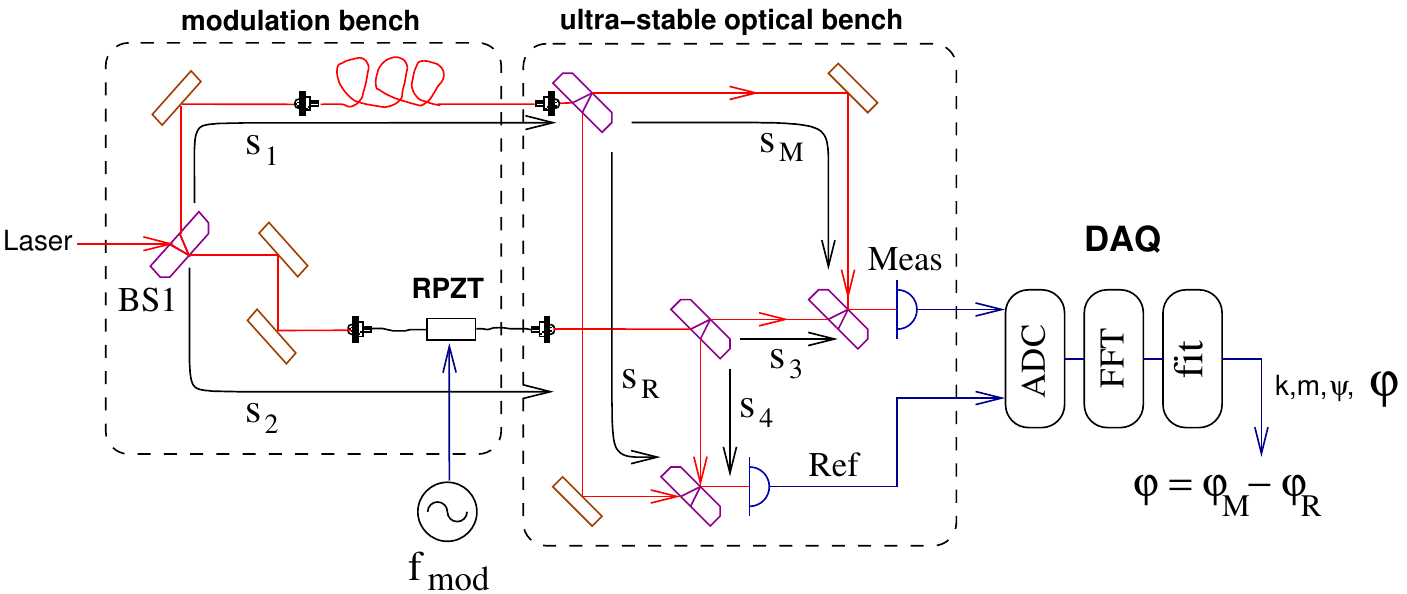}
\caption{Schematic overview of the experiment.}
\label{dpm_optosetup}
\end{center}
\end{figure*}

We have applied this technique to a very stable interferometer, namely the
engineering
model of the LISA Pathfinder (LPF) optical
bench\,\cite{ltp}, which consists of a 20\,cm$\times$20\,cm
Zerodur$^{\begin{scriptsize}\registered\end{scriptsize}}$ baseplate
with optical components fixed by hydroxide-catalysis bonding\,\cite{bonding}.
This
optical bench has been extensively characterized as part of ground testing
campaigns for the optical 
metrology of the LISA Pathfinder mission\,\cite{ltptests}, and its optical
pathlength stability has been measured to be better than 
5\,pm$/\sqrt{\rm Hz}$ above 1\,mHz. A non-planar ring oscillator (NPRO) Nd:YAG 
laser producing 300\,mW at 1064\,nm was used as light source.
For the experimental test, we chose a two-beam Mach-Zehnder interferometer,
using
self-assembled fiber-coupled phase modulators 
consisting of single-mode fiber optics coiled around ring piezo-electric
transducers (RPZT) in order to reach high modulation depths (up to 10 or
20).
Figure~\ref{dpm_optosetup} shows a schematic overview of 
the setup. 
The laser beam is split into two equal parts at the first beamsplitter BS1. A
RPZT driven by a sinusoidal voltage of approximately
4.5\,$\mathrm{V}_{\mathrm{pp}}$ at $f_{\rm mod}=280\,$Hz, 
produces a phase modulation of modulation depth $m\approx9.7$ in one of the two
beams. This portion of the optical setup 
denoted as modulation bench, contains the first beamsplitter BS1, phase
modulator, and corresponding fiber coupling devices 
which are all mounted on a standard metal optical breadboard. A single-mode
fiber feed-through is used to bring the main laser beam into
a vacuum chamber where both, the modulation bench and the optical bench reside.
The LISA Pathfinder optical bench is a set of 
four non-polarizing Mach-Zehnder interferometers, three of which have been used
in these experiments. The first one
--denoted \textbf{ \textit{M}}-- measures distance fluctuations between two
mirrors mounted on 3-axes piezo-electric
actuators\,\cite{ltpsubtraction}.
A second one --denoted \textbf{ \textit{R}}-- serves as phase reference to
cancel common-mode pathlength fluctuations that arise at the 
modulation bench, such as in metal mounts, phase modulator, and fiber optics.
The third interferometer --denoted \textbf{ \textit{F}}-- has an 
intentionally large optical pathlength difference of approximately 38\,cm, and
is used to measure laser frequency fluctuations. 
If we denote by $s_M$ and $s_R$ the optical pathlengths of the measurement and
reference interferometer, respectively, the 
phases emerging from the fit algorithm are given by
\begin{eqnarray}
\varphi_M &= \frac{2\,\pi}{\lambda} \left\{(s_1+s_M)-(s_2+s_3) \right\}= 
   \frac{2\,\pi}{\lambda} \left\{ (s_M-s_3)+\Delta \right\},\,\,\,\,\,\,\,\,\\
\varphi_R &= \frac{2\,\pi}{\lambda} \left\{ (s_1+s_R)-(s_2+s_4) \right\} = 
   \frac{2\,\pi}{\lambda} \left\{ (s_R-s_4)+\Delta \right\},\,\,\,\,\,\,\,\,
\end{eqnarray}
where $\lambda=1064$\,nm is the laser wavelength, $s_\mathrm{x}$ are the optical
paths outlined in Figure~\ref{dpm_optosetup}, and $\Delta=s_1-s_2$ represents
the common-mode pathlength difference 
between the two beams that includes everything starting from the first
beamsplitter BS1, the modulator, fiber optics up to the 
beamsplitters on the stable optical bench. Typically, the fluctuations of
$\Delta$ are several $\mu$m on 10\dots1000 second time scales 
and thus much larger than what we want to measure. However, the optical
pathlengths $s_R$, $s_3$ and $s_4$ are 
confined to the stable optical bench and have only negligible fluctuations. By
measuring both $\varphi_M$ and $\varphi_R$ and 
computing their difference
\begin{equation}
\varphi=\varphi_M-\varphi_R=
\frac{2\,\pi}{\lambda} \left\{ s_M-(s_R+s_3-s_4)\right\}
\label{cancel}
\end{equation}
it is possible to cancel the common-mode fluctuations $\Delta$ and to obtain a
measurement that is dominated by the fluctuations of
$s_M$ as desired. All photodetectors are indium gallium arsenide (InGaAs)
quadrant diodes with 5\,mm diameter. The 
photocurrent of each quadrant is converted to a voltage with a low-noise
transimpedance amplifier, filtered with a 9-pole 8\,kHz 
Tschebyscheff anti-aliasing filter and digitized at a rate
$f_\mathrm{samp}=20\,$kHz by a commercial 16-channel, 
16-bit analog-to-digital converter (ADC) card installed in a standard PC running
Linux. The time series are split in
segments of $N_{\mathrm{FFT}}=1000$ samples and transformed by a FFT
algorithm\,\cite{fftw}. The $N=10$
complex amplitudes of bins 1\dots10 of $f_\mathrm{mod}$ at frequencies
$280\dots2800$\,Hz are then fitted. This configuration allows us to reach a
real-time phase
measurement rate 
$f_\varphi=f_\mathrm{samp}/N_{\mathrm{FFT}}=20\,\mathrm{Hz}$.
\section{Noise investigations}\label{noise_inv}
During test and debugging experiments, two main noise sources were identified to
limit the interferometer sensitivity with this 
technique, which are laser frequency noise, and the frequency response (transfer
function) of the DAQ analog 
electronics, including photodiode transimpedance amplifiers and anti-aliasing
filters. In the following we explain the coupling 
mechanism of these noise sources, and the mitigation strategies we implemented
to counteract them.
\subsection{Laser frequency noise}\label{dpm_laserfreqnoise}
Laser frequency noise translates into phase readout noise in any interferometer,
whose pathlength difference $\Delta s$ 
between the two interfering beams is not exactly zero. In the case of the LPF
optical bench, this pathlength mismatch
has been determined to be approximately $10\,\mathrm{mm}$\,\cite{ltp}.  The
free-running frequency noise $\delta\nu$ 
of an unstabilized Nd:YAG NPRO laser at $10\,\mathrm{mHz}$ has been measured to
be of the order
of $2\times10^6\,\mathrm{Hz}/\sqrt{\mathrm{Hz}}$\,\cite{ghh:2004}. The
conversion
factor from laser frequency 
fluctuations $\delta\nu$ into phase fluctuations $\delta\varphi$ is given by the
difference in time of travel between the two beams
$\Delta s/c$, such that an estimate of the noise level can be calculated as
\begin{equation}
\delta\varphi=2\pi\frac{\Delta
s}{c}\delta\nu\approx2\pi\frac{\,10^{-2}\mathrm{m}}{3\times10^8\,\mathrm{m/s}}
2\times10^6\,\mathrm{Hz}=0.4\,\mathrm{mrad}/\sqrt{\mathrm{Hz}},
\label{dpm_deltanu}
\end{equation}
which limits the interferometer optical pathlength resolution $\delta s$ to
\begin{equation}
\delta
s=\frac{\lambda}{2\pi}\,\delta\varphi=\frac{1064\,\mathrm{nm}}{2\pi}\,0.4\,
\mathrm{mrad}/\sqrt{\mathrm{Hz}}=68\,\mathrm{pm}/\sqrt{\mathrm{Hz}}.
\label{dpm_delta_s}
\end{equation}
We implemented two mitigations strategies to correct for this effect and improve
the length resolution. Both 
methods worked similarly well and allowed suppression of this error below the
other noise terms. The first one is based on the
active laser frequency stabilization, for which we have used a commercial
iodine-stabilized Nd:YAG laser. The second method 
uses the third interferometer \textbf{\textit{F}} (mentioned above) to
independently measure a phase proportional to the amplified laser frequency fluctuations,
and applies a noise subtraction 
technique\,\cite{ltpsubtraction,felipe_phd,noisesub}
that properly estimates the coupling factor and removes the contribution from
the final data stream. The phase of this interferometer was read
out with the same deep modulation method as the main channels, and is dominated
by laser frequency fluctuations due to 
its large pathlength difference ($\approx 38\,$cm). A third method can also be
easily implemented as an active stabilization loop by
feeding back to the laser over a digital-to-analog converter the output of a
digital controller that uses the
difference phase extracted from interferometers \textbf{\textit{F}} and
\textbf{\textit{R}} as error signal.
\subsection{Frequency response of data acquisition
system}\label{dpm_optotfnoise}
The dominant error was identified to be the frequency response of the analog
electronics of the data acquisition system, 
in particular the contribution of the photodiode transimpedance amplifiers and
the anti-aliasing filters. The transfer function (TF) of
this analog portion of the DAQ shows small ripples in its magnitude of the order
of 0.9\,dB. The desired parameter $\varphi$
is essentially determined by running a fit onto the relative amplitudes of the
10 measured harmonic components 
$\tilde{\alpha}_n(m,\varphi)$. The ripples are, however, large enough to alter
the ratio between the harmonic amplitudes, such that the 
fit algorithm is disturbed, resulting in a high noise level.
We removed this error by separately measuring the transfer function of each
channel of the analog front end, fitting it to a 
model, and correcting accordingly the measured complex amplitudes
$\tilde{\alpha}_n(m,\varphi)$ before entering the fit routine. Thus,
we obtained the corresponding TF complex values 
$\beta_{\mathrm{n}}=b_{\mathrm{n}}\,\mathrm{e}^{i\theta_{\mathrm{n}}}$ for the 
10 frequency bins of interest $280-2800$\,Hz. Hence, the measured complex
amplitudes $\tilde{\alpha}_n(m,\varphi)$ were corrected as
\begin{equation}
\tilde{\alpha}'_{n}(m,\varphi)=\frac{\tilde{\alpha}_n(m,\varphi)}{\beta_n},
\label{dpm_tfcorrection}
\end{equation}
By using complex numbers, this correction also accounts for the TF phase shift,
and improves the estimation capability of the
modulation phase $\psi$.
\section{Optical length and attitude measurements}\label{dpm_lengthmeas}
\begin{figure}[bth]
\begin{center}
\includegraphics[width=\columnwidth]{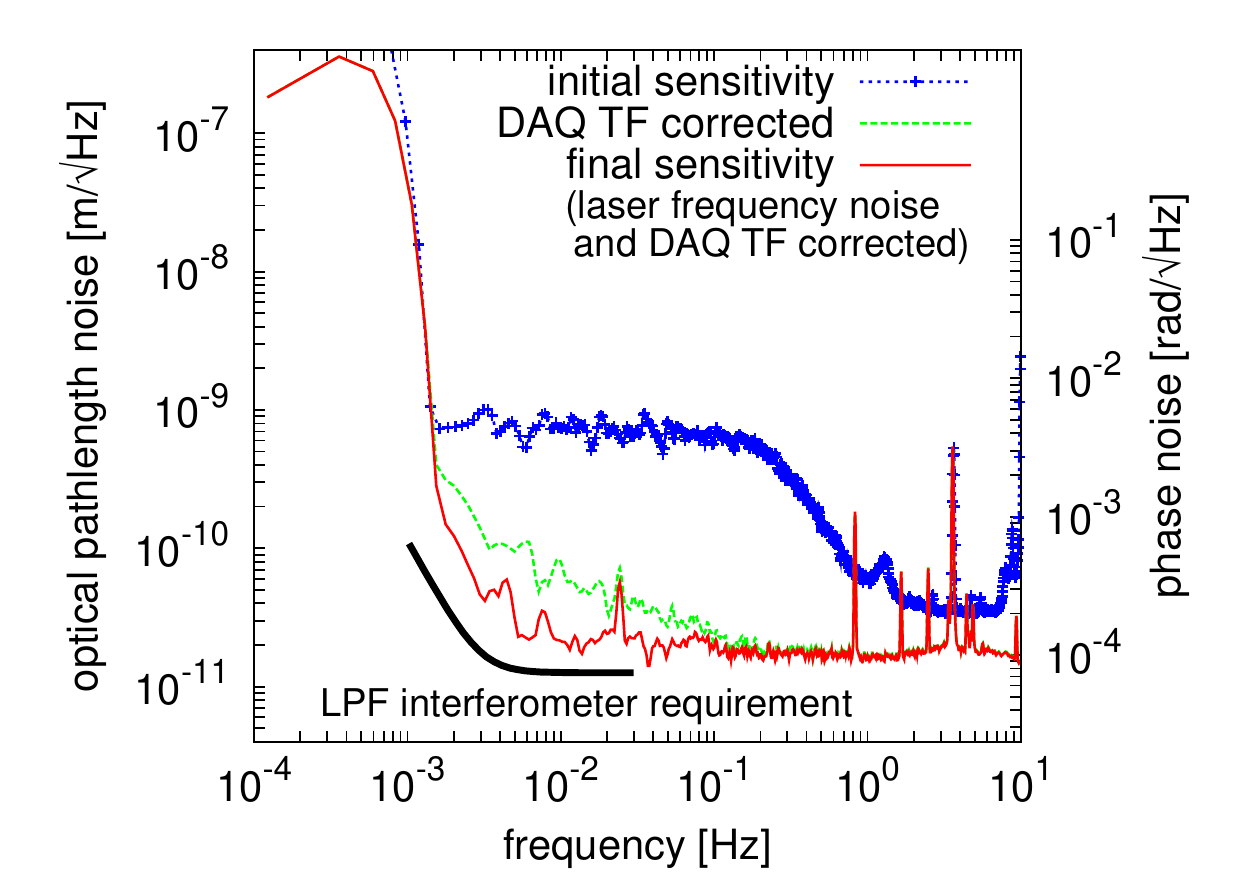}
\caption{Sensitivity of real optical pathlength measurements. Dashed curve with
crosses: initial sensitivity prior to
noise correction techniques. Dashed curve: sensitivity upon correction of DAQ
frequency response. 
Solid curve: sensitivity reach after application of noise mitigation strategies
-laser frequency noise and DAQ frequency response-.}
\label{dpm_length}
\end{center}
\end{figure}

\begin{figure}[bth]
\begin{center}
\includegraphics[width=\columnwidth]{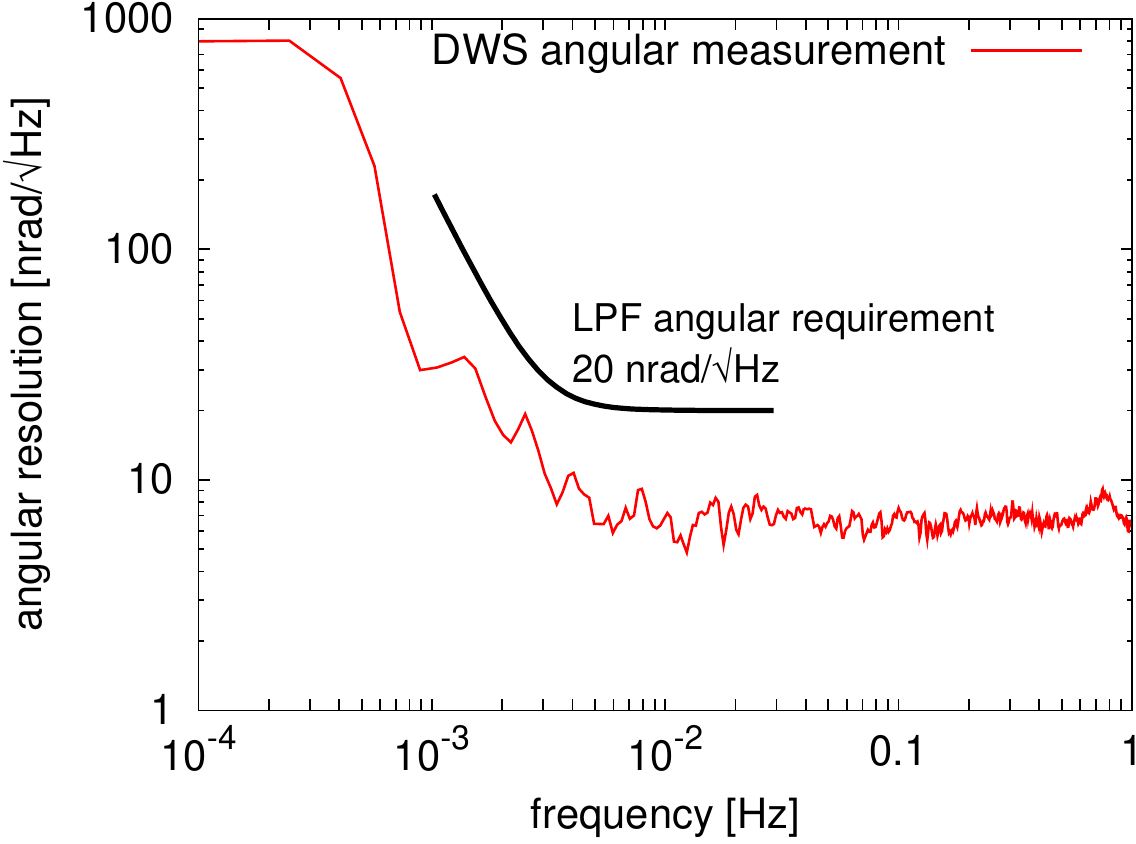}
\caption{Angular resolution obtained by applying a DWS algorithm to the phases 
extracted from individual cells of a quadrant photodetector.}
\label{dpm_dws}
\end{center}
\end{figure}

The experimental setup of Figure~\ref{dpm_optosetup} was used to conduct
long-term interferometric length
measurements on the LPF optical bench. Figure~\ref{dpm_length} shows the results
obtained in form of linear spectral
densities.
The dashed curve with crosses is the sensitivity obtained initially with this
method, without applying any 
of the noise mitigation strategies explained in Section~\ref{noise_inv}. The
dashed curve is the sensitivity 
achieved after applying the complex value correction of the DAQ frequency
response to the measured harmonic amplitudes
$\tilde{\alpha}_n(m,\varphi)$ (as given by Equation~\ref{dpm_tfcorrection}),
resulting in a sensitivity improvement of about 
one order of magnitude. The solid curve is the measurement length sensitivity
reached
upon subtraction of laser frequency noise, which increases the length resolution
in an additional factor of approximately 3.5 
at 10\,mHz. The measured optical pathlength sensitivity of this technique is of
the order of 
$20\,\mathrm{pm}/\sqrt{\mathrm{Hz}}$ above 3\,mHz and approximately a factor of
2 above the performance required
to the LPF interferometry, which has been plotted for comparison purposes.
As mentioned above, all photodetectors at the interferometer outputs on the
LPF optical bench are quadrant cell diodes.
The phases extracted from each individual quadrant cell are processed by a
differential wavefront sensing (DWS) 
algorithm\,\cite{morrison-1,morrison-2}, in order to measure the interferometer
alignment with high angular resolution.
The results of this measurement are shown on Figure~\ref{dpm_dws} as a linear
spectral density.
As it can be read from the plot, this technique reaches an angular sensitivity
better than 
$10\,\mathrm{nrad}/\sqrt{\mathrm{Hz}}$ above 3\,mHz, meeting with sufficient
margin the requirements set to the LPF
interferometry that have been also included in the graph as a comparison.
\section{Comparison with other techniques}
The only method known to the authors that allows length and angular
measurements at arbitrary operating points with low noise at millihertz frequencies is heterodyne
interferometry as described in Ref.~\cite{ltp}. The deep phase modulation method
presented here needs,
in comparison, much simpler beam generation hardware, namely one low-frequency
phase modulator like a piezo-electric transducer, as opposed to two AOMs with RF
driving electronics.
On the other hand, the data processing for phase extraction is more complicated,
which, however, becomes a smaller disadvantage with cheap processing power. The
heterodyne method typically requires additional stabilization
loops\,\cite{wand,ghh-ltpnoise} to reach
noise levels at $\mathrm{pm}/\sqrt{\mathrm{Hz}}$, e.g. for the laser power and
certain common-mode pathlengths (see Ref.~\cite{ltp}). The experiments described
above in Section~\ref{exp_setup}
show that these stabilizations are not required for the deep phase modulation
technique.
\section{Conclusions}
We have presented an interferometry technique for high sensitivity length and
angular optical measurements. This technique is based on the deep phase
modulation (over several radians) of one
interferometer arm and can be considered as an extension of the well-known
``$J_1\dots J_4$''
method\,\cite{su89,jin91,su93}. The harmonic amplitudes are used to numerically
solve an
overdimensioned system of equations to extract the interferometer phase and
other useful interferometer variables. This technique has been applied to
experiments
conducted 
on a very stable interferometer (the engineering model of the LISA Pathfinder
optical bench), achieving
an optical pathlength readout sensitivity of the order of
$20\,\mathrm{pm}/\sqrt{\mathrm{Hz}}$
(which translates to $10\,\mathrm{pm}/\sqrt{\mathrm{Hz}}$ for free-floating test
mass displacement), 
and alignment measurements with an angular resolution better than
$10\,\mathrm{nrad}/\sqrt{\mathrm{Hz}}$
in the millihertz frequency band. This performance is comparable to the best
heterodyne interferometers, and, e.g., only
a factor of 2 above the LISA Pathfinder pathlength measurement requirements. Two
main noise sources were identified, 
namely laser frequency fluctuations and the frequency response of the analog
portion of the data acquisition system, 
which both were completely mitigated by appropriate data processing methods,
hence improving the performance of
this technique by over a factor 35. Unlike other interferometry 
techniques, no additional control loops, for instance,
to actively stabilize the
optical pathlength difference or laser power fluctuations, have been
implemented 
or are required to reach the current sensitivity. Nonetheless, this could easily
be
done, in order to further improve the performance
of this method.
\section*{Acknowledgments}
We gratefully acknowledge support by the Deutsches Zentrum f\"ur Luft- und
Raumfahrt (DLR) 
(references 50 OQ 0501 and 50 OQ 0601).

\end{document}